\documentstyle[aps,prb,graphicx]{revtex}

\begin{document}
\twocolumn[\hsize\textwidth\columnwidth\hsize\csname
@twocolumnfalse\endcsname
\title{Novel Vortex Solid State in YBa$_2$Cu$_3$O$_{7 - \delta}$ Twinned Crystals.}
\author{B. Maiorov, and E. Osquiguil}
\address{Centro At\'{o}mico Bariloche and Instituto Balseiro, \\Comisi\'on Nacional de
Energ\'{\i}a At\'omica, 8400 S. C. de Bariloche, R. N., Argentina.}
\date{\today}
\maketitle
\begin{abstract}
We report on the scaling of transport properties around the vortex melting in
YBa$_2$Cu$_3$O$_{7- \delta}$ oriented-twin single crystals in applied magnetic fields 
between 1T and 18T. We find that for all the measured field range the linear
resistivity scales as $\rho (t,\theta) \sim t^{sy} {\cal F}_{\pm} (\sin(\theta)t^{-sx})$,
with $t=|T-T_{BG}|$ and $\theta$ the angle between de planar defects and the 
magnetic field. The scaling is valid only for angles where the transition
temperature $T_{BG} (\theta)$ shows a cusp. The critical exponents $sx$ and $sy$ are  in
agreement with the values predicted by Lidmar and Wallin only at magnetic fields
below 4T. A change in the value of $sx$ from $sx = 1 \pm 0.2$ to $sx = 3 \pm 0.2$ at around
$H^{cr} \approx $ 4T when the magnetic field is increased, is responsible for changes in the
shape of the $T_{BG} (\theta)$ curve and in the dependence of the linear dissipation on
temperature and angle. The results strongly suggest the existence of a different vortex
glassy phase in twinned crystals compared to the Bose-glass state found in samples
with linear defects.
\end{abstract}

\vskip1pc] \narrowtext
One of the most important aspects in high temperature superconductors is the existence of a
melting line in the magnetic field-temperature $(H-T)$ vortex phase diagram. The quest to
understand the thermodynamic nature of the transition has been reflected in an important
number of publications.\cite{algunreview} Nowadays it is accepted that in clean crystals
the solid-liquid transition is first order, while the introduction of random point-like or
correlated disorder transforms it into second order. Due
to the high value of the Ginsburg parameter in high $T_c$ materials, the critical region
associated with this kind of transition becomes experimentally accessible. This important
fact has led to the development of scaling theories
\cite{vortexglass,boseglass} which describe the vortex dynamics near the transition,
enclosing its properties into two critical exponents. In both theories the critical
exponents settle the way in which, the correlation length, $\xi \sim |T - T_{BG}|^{-\nu} =t^{- \nu}$,
and the time scale in which a fluctuation of this size survives, $\tau \sim \xi^{z}$, 
diverge as the transition is approached. In the Bose-glass theory {\it linear} correlated
defects induce an anisotropic correlation volume \cite{boseglass} with $\xi_{\parallel} \sim \xi_{\perp}^2$
 which are, respectively, the correlation lengths parallel and
perpendicular to the correlated defects. In this theory \cite{boseglass} the linear
resistivity is predicted to scale as
\begin{equation}
\label{scalgral}
\rho (t,\theta) \sim t^{sy} {\cal F}_{\pm} (\sin(\theta) t^{-sx})
\end{equation}
with a phase boundary separating the Bose-glass and the vortex liquid of the form
\begin{equation}
\label{scalpico}
T_{BG}(\theta) = T_{BG}(0)[1-(\frac{\sin(\theta)}{x_c})^{1/sx}]
\end{equation}
where $sy = \nu (z-2)$, $sx = 3 \nu$, and $x_c = \sin(\theta_c)  t^{sx}$.
Lidmar and Wallin pointed out \cite{lidmar} that the 
correct thermodynamic variable to be included in the scaling anzats for the
dynamical response of the vortex system in tilted magnetic fields is {\bf B}, instead of 
{\bf H} as used in ref. \cite{boseglass}.
This important remark leads to a modification of eqs. \ref{scalgral} and \ref{scalpico}
through a change in the dependence of $sx$ on the critical exponent $\nu$.\cite{lidmar}, \cite{nelson} Accordingly, the form of the new phase boundary is given by 
eq. \ref{scalpico} with $sx = \nu$, while the linear resistivity still scales as 
eq. \ref{scalgral} but again with $sx = \nu$.

Equation \ref{scalpico} describes one of the key characteristics of the Bose-glass to
liquid transition, {\it i.e.} the existence of a cusp in the transition curve at $\theta =
0$. This cusp was first observed by Worthington {\it et al.} \cite{worthington} in twinned
YBa$_{2}$Cu$_{3}$O$_{7-\delta}$ crystals. Fleshler {\it et al.} \cite{fleshler}
studied the vortex dynamics for different orientations of the twin
boundaries with respect to the applied field and current, and explained the cusp in terms of pinning effects by the twin potentials. The relation between
the cusp and the Bose-glass transition was shown by Jiang {\it et al.} \cite{jiang}
through transport measurements in YBa$_{2}$Cu$_{3}$O$_{7-\delta}$ crystals with
columnar defects. Later on, Grigera {\it et al.}
\cite{grigera} identified a Bose-glass phase in twinned YBa$_{2}$Cu$_{3}$O$_{7-\delta}$
crystals through a scaling analysis of the vortex dynamics near the
cusp line. An interesting point related to the shape of the cusp, however, has gone unnoticed in the
literature. While in samples with columnar defects the transition curve is well
described by $T_{BG} (\theta) \sim 1 - |\theta |$ at small angles, \cite{jiang,klein,smith}
in twinned crystals   a non-linear transition curve
\cite{worthington,fleshler,grigera,safar,morre} is observed. Monte Carlo simulations of the
Bose-glass melting \cite{lidmar} in a vortex system interacting with linear
  defects gave $\nu \approx 1$ which, through eq. \ref{scalpico}, seems to account
only for the results in samples with columnar defects. It is therefore important to clarify
whether the cusp in samples with planar correlated defects arises as a consequence of a
liquid to Bose-glass transition as suggested in ref. \cite{grigera}, and to
which extent the scaling theory describing the vortex dynamics is applicable to this more 
anisotropic case.

In this paper we show that within the linear regime the scaling hypothesis
$\rho (t,\theta) \sim t^{sy} {\cal F}_{\pm} (\sin(\theta) t^{-sx})$ 
is fulfilled for all magnetic fields between 1T and 18T in 
YBa$_{2}$Cu$_{3}$O$_{7-\delta}$ heavily oriented-twin single crystals. For all applied
fields the transition temperature $T_{BG}$ shows a cusp
around $\theta = 0$. However the shape of the $T_{BG}(\theta)$ curve is only consistent
with the predictions of Lidmar and Wallin for $H < H^{cr} \approx 4T$. Above this field a
change of the cusp's shape is accompanied by a sudden increase in $sx$ from $sx = 1 \pm 0.1$ to $sx = 3 \pm 0.2$ while $sy$ remains 
constant in the whole field range. This jump of $sx$ is clearly reflected in the scaling of the linear resistivity. We discuss these
findings in terms of a change in the universality class of the vortex solid-liquid
transition which we adscribe to a corresponding change in the nature of
the vortex solid phase as the field is swept through $H^{cr}$.

\begin{figure}
\centering
\includegraphics[width=86mm]{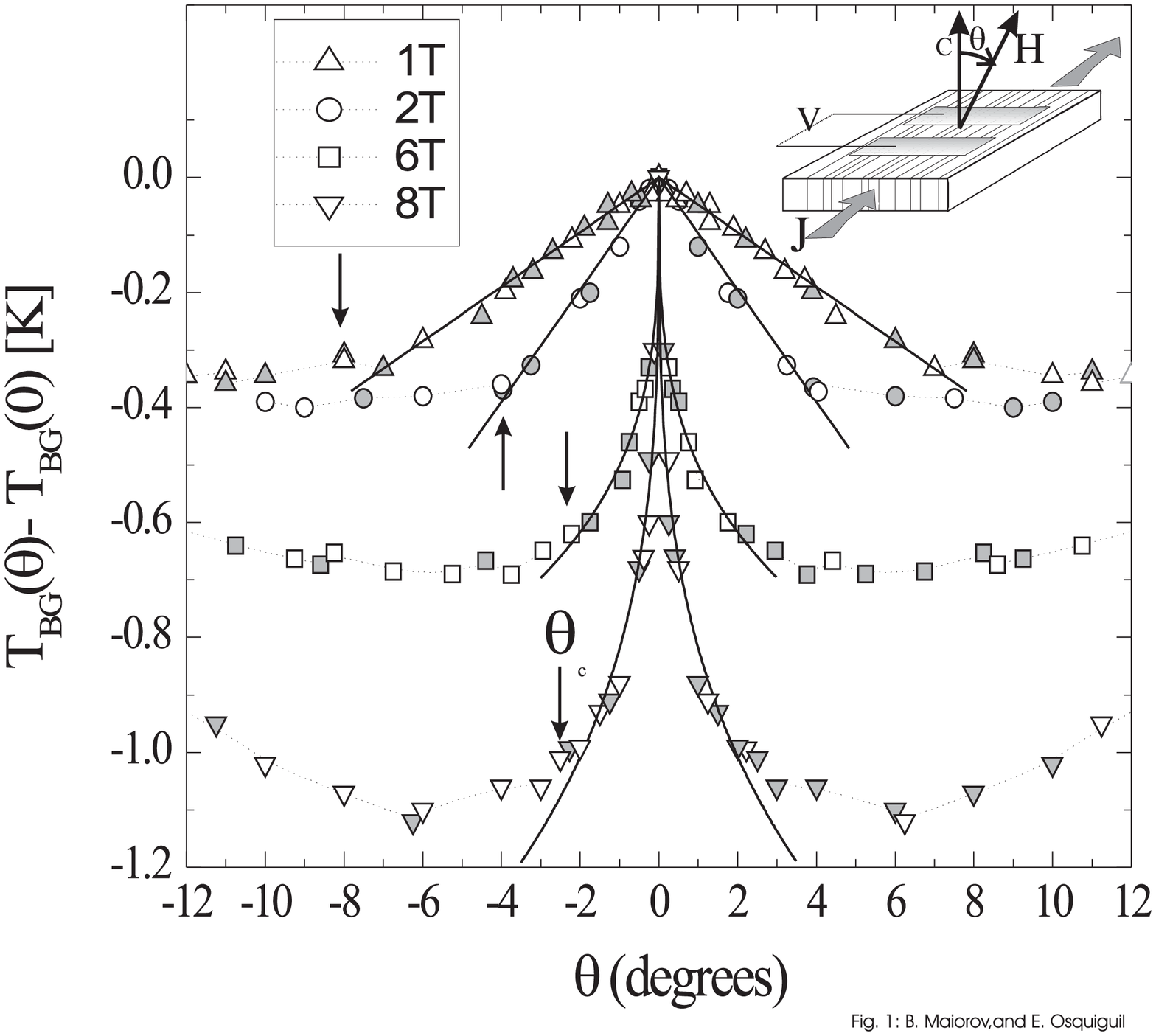}
\caption[]{Angular dependence of $T_{BG}(\theta)-T_{BG}(0)$. Solid lines are fits to eq. \ref{scalpico} with $sx = 1$ for data at 1T and 2T, and $sx=3$ for
data at 6 and 8T. The arrows indicate  the critical angle $\theta_c$ above which the linear resistivity does not scale. Data point are mirrored, gray points correspond to mirrored data. Inset: experimental setup for transport measurements.}  
\label{figura1}
\end{figure}

The YBa$_{2}$Cu$_{3}$O$_{7-\delta}$  crystal with oriented twins was mounted onto a rotatable 
sample holder with an angular resolution of $\Delta \theta \approx 0.05^{\circ}$
inside a cryostat provided with an 18T magnet. The crystal has a critical temperature of $T_c\sim 93$K and $\Delta T_c\sim 0.5$K width dimension of $1 \times 0.5 \times 0.03$mm$^3$. A uniform $dc$-current was
injected through the crystal parallel to the twin planes and the crystal
was rotated at angles $\theta$ in a plane perpendicular to the twins 
in a constant Lorenzt force configuration (see inset Fig. \ref{figura1}). In this 
setup the vortices are moved and tilted in a direction perpendicular to the 
defects. All transport measurements were performed within the linear regime at J=$3$Acm$^{-2}$. 

In Fig. \ref{figura1} we show the transition temperature $T_{BG} (\theta)$ for different 
magnetic fields. This temperature has been defined from $\rho (\theta, T)$ measurements as
the temperature at which the data falls below our experimental resolution ($V/I \leq 20 \mu\Omega$).  The value at $\theta=0$ coincide within experimental error with that obtained from the scaling of the linear resistivity. It is seen that the shape of the cusp
changes as the magnetic field is increased. The change is rather
abrupt, and occurs at a crossover field $H^{cr} \approx 4T$. Below $H^{cr}$ the $T_{BG}
(\theta)$ is linear in $|\theta|$, while above a non-linear dependence develops. The solid lines, which are fits to eq. \ref{scalpico} with $sx
= 1$ for H= 1T and 2T, and $sx = 3$ for 6T and 8T illustrate this point. The
arrows  correspond to a critical angle, $\theta_c$, above which the scaling properties of
the linear resistivity are lost.  Not only the shape of the cusp changes, but
its amplitude increases with field. The angular width of the cusp shrinks when the
field increases but remains almost constant for $H \geq 4$T. Interestingly enough, the cusp
exists up to the highest magnetic fields we measured (18T).
This may be compared to the behavior reported for $T_{BG} (\theta)$ in (Ba,K)BiO$_3$
irradiated cubic crystals, \cite{klein} where the cusp gradually disappears upon
surpassing the matching field, $B_{\phi}$, in which the number of vortices equals the
number of the columnar defects. The absence of saturation   in
our experiments indicates that the twin boundary potentials are still efficient to induce
vortex confinement and localization in a field range in which induced columnar tracks with
  $B_{\phi} \approx 18T$ would destroy the superconducting properties of the crystal.

\begin{figure}
\centering
\includegraphics[width=86mm]{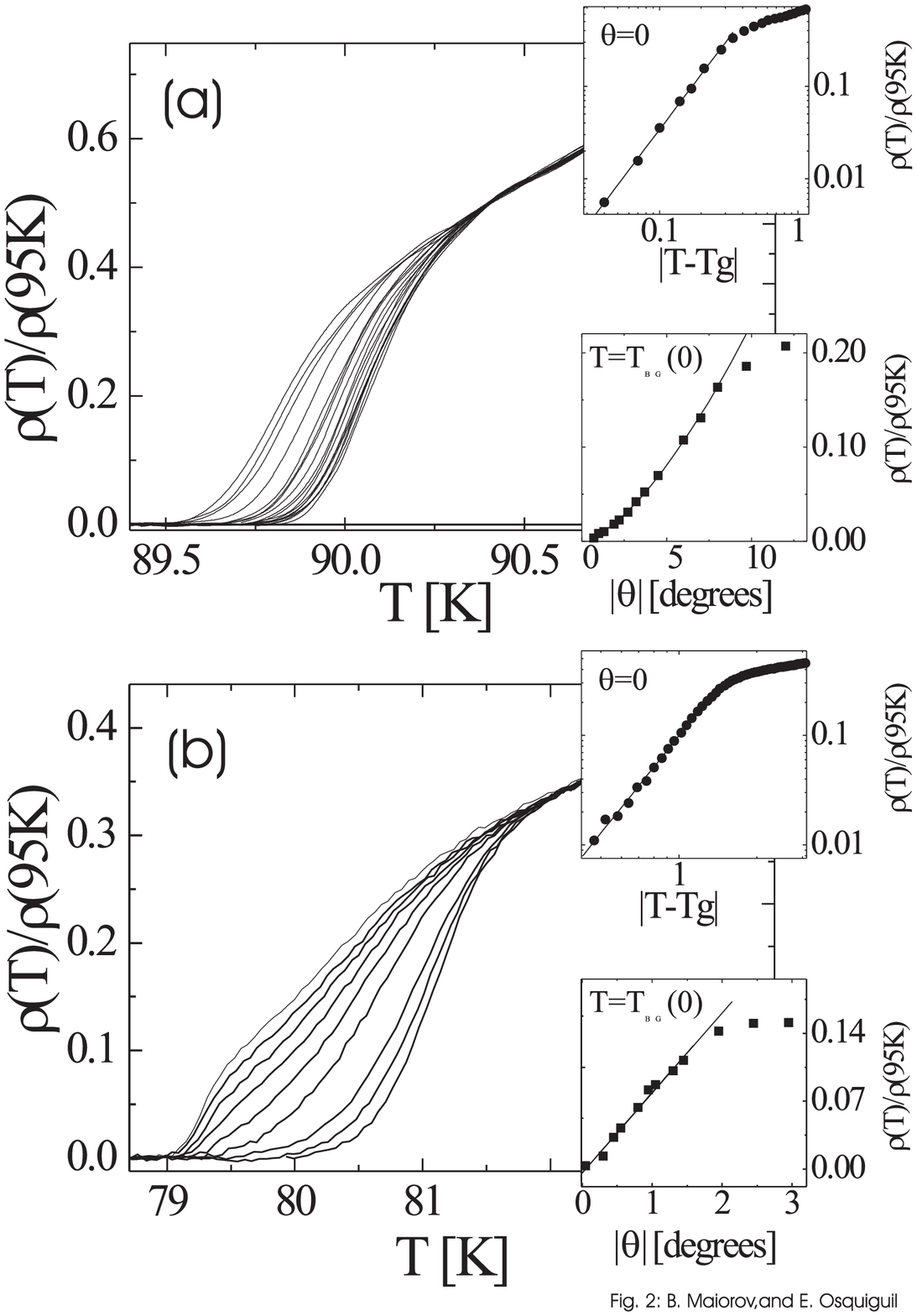}
\caption[]{{\bf (a)} Normalized resistivity $\tilde\rho=\rho/\rho(95K)$ for different tilting angles up to $\theta_c$ at H=1T, $\theta = $0.2, 0.45, 0.65, 1.0, 1.3, -1.3, 1.6, 1.9, 2.2, 2.7, 3.2, 3.7, -4.5, 6.0,-7.0, 8.0, -8.0 deg. 
Upper inset: $\tilde\rho(t,0)$ . The solid line is a fit using $\tilde\rho \sim t^{sy}$ with 
$T_{BG}(0)=(89.82\pm 0.05)$K and $sy=2.5\pm 0.3$. Lower inset: $\tilde\rho(0,\theta)$. The solid line is a fit to $\tilde\rho\sim\theta^{sy/sx}$ with $sy/sx=2.2 \pm 0.3$. {\bf (b)} idem for H=9T, $\theta =$-0.15, 0.3, 0.75, 0.85, 1.1, 1.25, 1.47, -1.8, 2.25 degrees. Upper inset:  $T_{BG}(0)=(79.85 \pm 0.1)$K and $sy=2.8\pm0.3$. Lower inset:  $sy/sx=1.1 \pm 0.3$.} 
\label{figura2}
\end{figure}

\begin{figure}
\centering
\includegraphics[width=86mm]{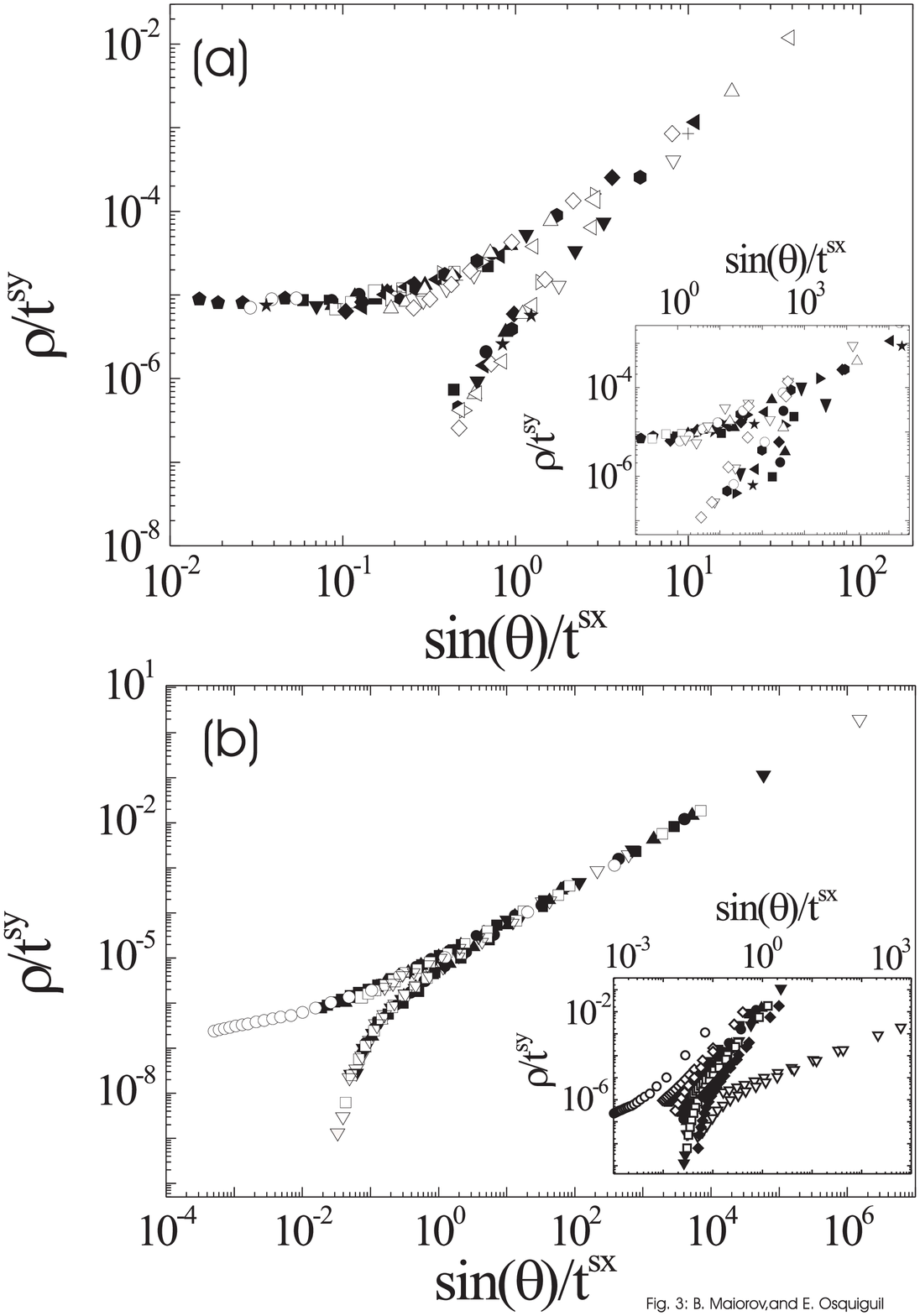}
\caption[]{ {\bf (a)} Scaling of the resistivity curves shown in figure \ref{figura2}(a) 
according to eq. \ref{scalgral} for $sy=2.3$, $sx=1.1$ and $T_{BG}=89.82$K. The inset shows 
the same scaling using $sx = 3$. {\bf (b)} Scaling of the resistivity curves shown in 
figure \ref{figura2}(b) according to eq. \ref{scalgral} for $sy=2.8$, $sx=3.0$ and 
$T_{BG}=79.85$K. The inset shows the same scaling using $sx = 1$.} 
\label{figura3}
\end{figure}

The changes in the shape of the phase boundary are also reflected in the dissipation in the
liquid state as shown in Fig. \ref{figura2}, in which $\rho(T)$ is plotted for different angles near $\theta = 0$. Panel (a) shows data characteristic of the low field regime, while panel (b) displays the data corresponding to the high field
region. It is worth noticing that the linear dissipation has a different
behavior at low and high fields as the angle between the magnetic field and the defects is
increased. At low $H$ the $\rho (T)$ curves shift almost parallel to each other  while at high fields $\rho (T)$ stretches to lower temperatures as $\theta$ is increased. These differences are indicative of a change in the vortex
dynamics below and above a crossover field $H^{cr}$ which separates the low and high field
regions. In order to quantify the observed changes we refer to the scaling theory
\cite{lidmar,nelson} according to which the linear resistivity, when the applied field is
parallel and vortices move perpendicular to the correlated disorder, should
scale as $\rho (T,0) \sim t^{sy}$ with $sy = \nu (z-2)$. On the other hand, the resistivity
as a function of the tilting angle at the transition temperature should behave as $\rho (0,\theta) \sim \theta^{sy/sx}$ for small angles. These scalings are shown in the insets of
panels (a) and (b) for the corresponding fields. In general we find that for $H < H^{cr}$,
$sy =2.4\pm 0.3$ and $sy/sx =2.3\pm 0.2$, while for $H > H^{cr}$  $sy =2.7\pm 0.3$ and $sy/sx =0.9\pm 0.2$. Note that within experimental error $sy$
seems to be field independent while the ratio $sy/sx$ undergoes an abrupt change. From these we conclude that the changes  in the behavior
of the linear resistivity are solely due to a change in the exponent $sx$.
In order to see this, we show in Fig. \ref{figura3} the scaled linear resistivity as
predicted by eq. \ref{scalgral}. Panel (a) shows the same data displayed in Fig. \ref{figura2}(a),
while in panel (b) the scaling of the data in Fig. \ref{figura2}(b) is shown. Both scalings are
very good for $\rho (T,\theta)$ curves below the corresponding critical angle
$\theta_c$. Curves for $\theta > \theta_c$ do not scale. An important point to note here is
that the critical exponent $sx$ is different at low and high fields. In
panel (a), $sx = 1 \pm 0.2$ was used to obtain a good scaling, while in panel (b) the
value of $sx$ was increased to $3\pm 0.2$ in order to make the curves to scale.  In the insets the same scaling is shown but forcing the
value of $sx$ to be 3 at low fields and 1 at high fields, and maintaining the rest of the
parameters constant, as required by consistency with eq. \ref{scalgral} at $\theta = 0$.
Clearly, this assignment of critical exponent values renders the scaling impracticable in
both cases, indicating that the change in the value $sx$ is outside the errors
involved in such scaling.  It is worth
noticing that an exponent $sx = 1$ corresponds, through eq. \ref{scalpico}, to a linear
$T_{BG} (\theta)$ dependence, which indeed is seen in Fig. \ref{figura1} for low fields, while a value
of $sx=3$ gives a non-linear shape as observed in Fig. \ref{figura1} at high fields.

\begin{figure}
\centering
\includegraphics[width=86mm]{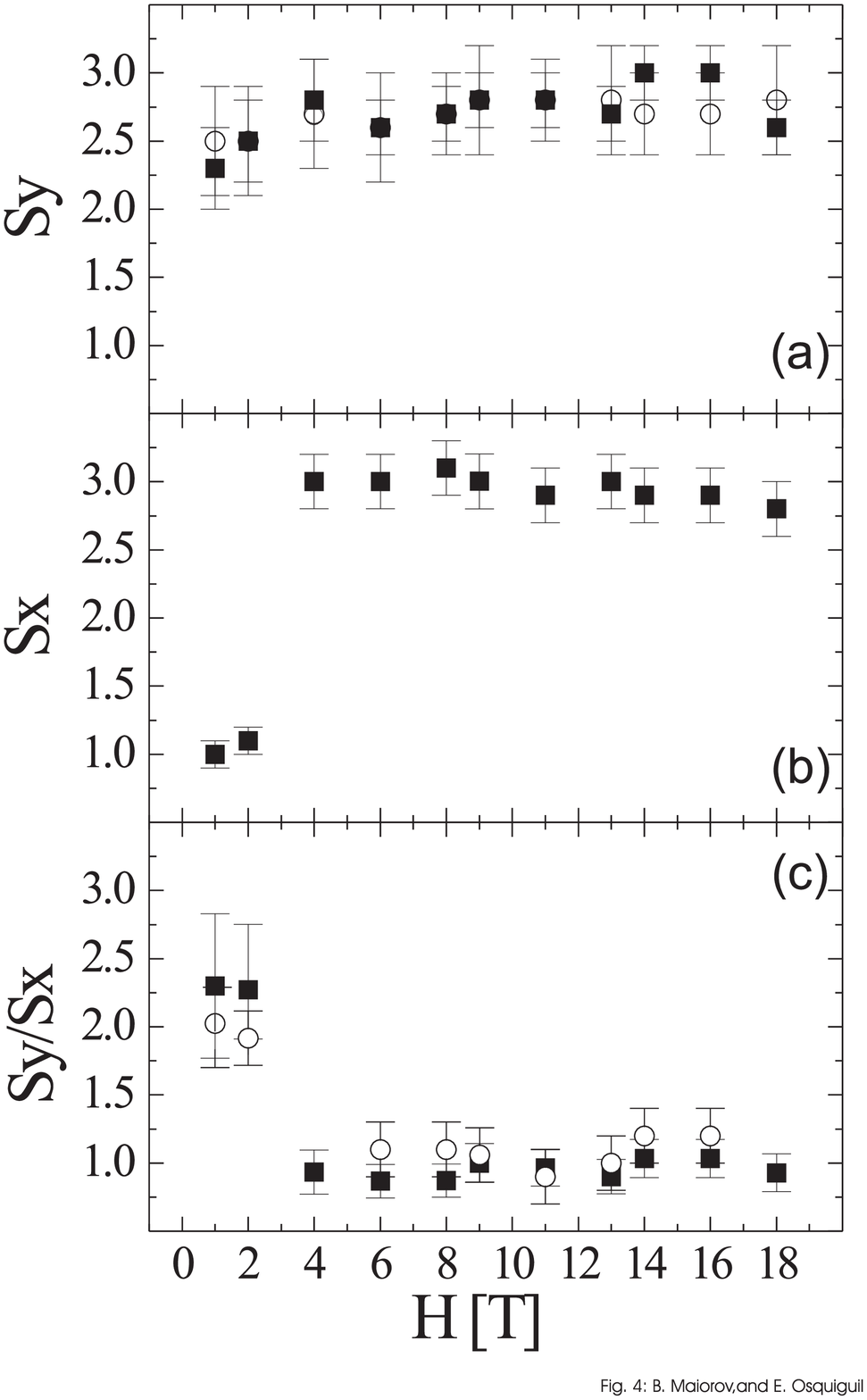}
\caption[]{Summary of the results of the scalings for different fields. {\bf(a)} Values of $sy$
according to the resistivity scaling (solid squares) and obtained from $\tilde\rho(t,0)\sim t^{sy}$
(open circles). {\bf(b)} Values of $sx$ according to the resistivity scaling. {\bf(c)} values
of $sy/sx$ from the fitting of $\tilde\rho(0,\theta)$ (open circles) and the ratio of $sy$ and
$sx$ obtained from the solid squares in panels (a) and (b).}
\label{figura4}
\end{figure}

In Fig. \ref{figura4} we show a summary of our findings by plotting the values of the two critical exponents and their ratio as a function of the magnetic field. The open circles in panel (a) correspond to the values of $sy$ obtained by a fit
of the resistivity data to $\rho(t,0)\sim t^{sy}$, from which also $T_{BG}(0)$ was
obtained. With these values of $T_{BG}(0)$ the scaling of $\rho(t,\theta)$, as
required by eq. \ref{scalgral}, was performed to obtain the solid squares in panel (a) and
(b). The error bars were estimated by the effect the variation of the exponents $sx$
and $sy$ has on the scaling of the linear resistivity. As is seen $sy$ is,
within error bars, independent of the magnetic field and of the way in which
it is obtained. The value of $sx$, however, jumps from 1 to 3 at 4T, remaining
constant for higher or lower fields. Finally in panel (c) we show the ratio of $sy/sx$.
The solid squares were calculated from the values in panel (a) and (b) while the open circles were obtained by adjusting the resistivity data at $T_{BG}(0)$ to $\rho(0,\theta)\sim \theta^{sy/sx}$. Note that the change at around 4T is independent of
the way in which $sy/sx$ was obtained.

An important conclusion that can be drawn from our results is that the way in which the
linear dissipation in the vortex liquid varies as a function of angle and temperature can
be accounted for by a scaling law of the form of eq. \ref{scalgral} independently of the
applied magnetic field. This is a fingerprint of a second order
vortex solid-liquid phase transition occurring at $T_{BG} (\theta)$ for angles below
$\theta_{c}$. Within this physical picture, our results indicate the
existence of two different field regimes characterized each one  by different values of the
critical exponents which govern the shape of the transition curve and the
dynamic response of the liquid phase in the linear regime. On general grounds the two
regimes can be interpreted in terms of a change in the universality class to which the
solid-liquid transition belongs. Let us first discuss the low field regime.

Our results for $H<H^{cr}$ closely resemble those observed in crystals with
columnar defects. We may therefore assume that the $T_{BG}(\theta)$ is linear in $|\theta|$ as a consequence of the interaction of the vortex system with some kind of columnar-type potentials. 
The identification of the columnar-type 
defects which might be present in our twinned crystal is however difficult. A possible candidate we found by optical inspection of our crystal were end points of twin planes located between the voltage contacts. An end point breaks the translational symmetry of the
twin plane and may emulate a columnar-type defect for vortices located near it. This end point corresponds microscopically to zones where a twin turns into a transverse one, or more likely in a microtwinned region which can not be seen under an optical microscope. 
Other source of columnar-type defects
which can also contribute to localise the vortices in this regime, might be screw 
dislocations which also exist in twinned YBa$_{2}$Cu$_{3}$O$_{7-\delta}$ crystals.
\cite{sun} Within this assumption a matching field should exist, above which all linear 
defects become occupied by a vortex line. In view of the work of Klein {\it et al.} showing 
that the cusp in $T_{BG} (\theta)$ exists up to magnetic fields more than three times larger 
than the corresponding $B_{\phi}$,
it is not straightforward in our case to identify the crossover field $H^{cr}$ with the 
matching field of the linear defects. It could well be that $H^{cr}$ represents the field
above which the effects of the columnar defects are completely washed out. In this case $H^{cr}$ should be sample dependent and could even be absent in crystals of very 
high quality. This physical picture could be tested in a more direct way
by measuring the vortex transport properties in an oriented-twin crystal with a low concentration
of columnar tracks induced by ion irradiation in the $c$-axis direction.

Once the effect of the columnar potentials has been saturated, the dynamics of the vortex
system is determined by its interaction with the twin planes. This corresponds to the field 
regime above $H^{cr}$. In this field range vortices localise inside the twin 
potentials, at $T_{BG} (\theta)$. 
However, the vortex solid that is formed above $H^{cr}$ ought to be different from that present 
in samples with columnar tracks alone. A way to see this is to consider the form in which
vortices can accommodate inside the correlated potentials. While columnar tracks allow single 
vortex occupancy only, in a finite sample with twin planes vortices will continue to accommodate
inside the correlated defects until the increase in energy due to their mutual interaction 
overcomes the energy gain due to the vortex-defect interaction. This difference in the way in which vortices interact with and become localised at the twin potentials gives rise to
a different dynamical response of the vortex liquid compared to that found in a system
interacting with columnar tracks. It is therefore reasonable to conclude that the vortex
liquid will freeze in a vortex solid which does not have the same structure as a
Bose-glass. The non-linear phase boundary in $\theta$ 
observed in our experiments may be taken as a fingerprint of this novel vortex solid phase.
In general we may say that the form of the phase boundary separating the solid and 
liquid phases is determined by the type of correlated disorder which is present in the sample. 

In summary, we have shown that the vortex solid to liquid transition in YBa$_{2}$Cu$_{3}
$O$_{7-\delta}$ oriented-twin crystals is second order within 1T and 18T, and occurs at a
phase boundary which is non-linear in angle when the vortex dynamics is dominated by the
interaction with planar defects. The linear dissipation in the liquid state in tilted
magnetic fields can be very well described by a scaling law theoretically proposed to
explain the dynamics of a vortex system interacting with columnar-type pinning
potentials, but with different critical exponents. These observations strongly indicate the
existence of a distinctly different vortex solid phase in twinned crystals compared to the
Bose-glass phase found in samples with columnar defects.

We would like to thank C. Balseiro, and F. de la Cruz for useful comments and discussions.
We also thank G. Nieva for providing the oriented-twin crystal and for bringing our
attention to ref. \cite{sun}. Work partially supported by grants from CONICET PIP 4207, and
ANPCyT protect PICT97-03-00061-01116. E. O. is member of CONICET, Argentina.

\noindent

\noindent

\noindent

\noindent

\end{document}